\documentclass[reprint,amssymb, amsmath, aps, superscriptaddress, showpacs, footinbib,  prb]{revtex4-1}
\usepackage{graphicx}
\pdfoutput=1
\usepackage{multirow}
\usepackage{color}

\newcommand{\eff}{\text{eff}}
\newcommand{\AFM}{\text{AFM}}
\newcommand{\FM}{\text{FM}}
\newcommand{\Done}{\text{D1}}
\newcommand{\Dtwo}{\text{D2}}

\newcommand{\iso}{\text{iso}}

\newcommand{\ccls}{Cu$_3$(AsO$_4$)(OH)$_3$}

\begin{document}

\title{Two energy scales of spin dimers in clinoclase Cu$_3$(AsO$_4$)(OH)$_3$}
\author{Stefan Lebernegg}
\affiliation{Max Planck Institute for Chemical Physics of Solids, N\"{o}thnitzer
Str. 40, 01187 Dresden, Germany}

\author{Alexander A. Tsirlin}
\email{altsirlin@gmail.com}
\affiliation{Max Planck Institute for Chemical Physics of Solids, N\"{o}thnitzer
Str. 40, 01187 Dresden, Germany}
\affiliation{National Institute of Chemical Physics and Biophysics, 12618 Tallinn, Estonia}

\author{Oleg Janson}
\affiliation{Max Planck Institute for Chemical Physics of Solids, N\"{o}thnitzer
Str. 40, 01187 Dresden, Germany}
\affiliation{National Institute of Chemical Physics and Biophysics, 12618 Tallinn, Estonia}

\author{Helge Rosner}
\email{Helge.Rosner@cpfs.mpg.de}
\affiliation{Max Planck Institute for Chemical Physics of Solids, N\"{o}thnitzer
Str. 40, 01187 Dresden, Germany}

\begin{abstract} Magnetic susceptibility and microscopic magnetic model of the
mineral clinoclase Cu$_3$(AsO$_4$)(OH)$_3$ are reported. This material can be
well described as a combination of two nonequivalent spin dimers with the
sizable magnetic couplings of $J\simeq 700$~K and $J_{\Dtwo}\simeq 300$~K.
Based on density functional theory calculations, we pinpoint the location of
dimers in the crystal structure. Surprisingly, the largest coupling operates
between the structural Cu$_2$O$_6$ dimers. We investigate magnetostructural correlations in Cu$_2$O$_6$
structural dimers, by considering the influence of the hydrogen position on the
magnetic coupling. Additionally, we establish the hydrogen positions that were
not known so far and analyze the pattern of hydrogen bonding.
\end{abstract}

\maketitle

\section{Introduction}
The majority of magnetic systems have more than one characteristic energy
scale, according to the different nature of interactions between the spins. For
example, in low-dimensional magnets strong interactions within chains or planes
are of direct exchange or superexchange type, whereas weak interchain
(interplane) couplings may have purely dipolar
origin.\cite{birgeneau1970,epstein1970,zhou2007} The different energy scale of
these interactions implies that, at sufficiently high temperatures, the
magnetic behavior is solely determined by the strong couplings, and the system
can be fully described in terms of a low-dimensional spin
model.\cite{birgeneau1970,[{For example: }][{}]ronnow1999,*goddard2012}
However, at low temperatures, interchain (interplane) couplings come into play,
and a full three-dimensional description is required.

In complex spin systems, the identification of different energy scales is by no
means a simple problem. Naively, one could think that distinct crystallographic
positions of magnetic atoms should lead to different strengths of magnetic
couplings and, therefore, to different energy scales of the magnetic behavior.
Indeed, in spin-dimer systems, such as BaCuSi$_2$O$_6$ and NH$_4$CuCl$_3$, the
spin-$\frac12$ Cu$^{2+}$ ions occupy several nonequivalent
positions\cite{sheptyakov2012,*ruegg2007,*kraemer2007,ruegg2004} and form
different types of spin dimers which have large impact on the high-field
behavior, including the unusual critical regime of the Bose-Einstein
condensation of magnons in BaCuSi$_2$O$_6$
(Refs.~\onlinecite{sebastian2006,*vojta2007,*laflorencie2011}) and the
fractional magnetization plateaus in NH$_4$CuCl$_3$
(Refs.~\onlinecite{shiramura1998,inoue2009}). However, spin dimers can also be
formed between two nonequivalent Cu positions, thus leading to only one type of
spin dimer and one energy scale, as in the spin-ladder compound
BiCu$_2$PO$_6$.\cite{koteswararao2007,*mentre2009,*tsirlin2010}

When the system contains several Cu$^{2+}$ positions with dissimilar local
environment and variable connectivity of the Cu polyhedra, the identification
of relevant interactions and energy scales becomes increasingly complex. The
problem of magnetic dimers that do not match structural
dimers,\cite{deisenhofer2006,janson2011,schmitt2010} as well as magnetic chains
running perpendicular to the structural
chains,\cite{kaul2003,*tsirlin2011,*janson2011b} is well-known in quantum
magnets and requires a careful microscopic analysis. 

Here, we report on the magnetic behavior and microscopic modeling of
clinoclase, Cu$_3$(AsO$_4$)(OH)$_3$. The intricate crystal structure of this
mineral\cite{ghose1965,eby1990} features three nonequivalent Cu positions.
Nevertheless, the resulting spin lattice comprises only two types of magnetic
dimers with notably different interaction energies. Our microscopic analysis
shows that one spin dimer is formed between two different Cu positions
(Cu1--Cu2) and does not match the respective structural dimer. However, the
other spin dimer coincides with the Cu3--Cu3 structural dimer. We argue that
neither a straightforward comparison of Cu--Cu distances nor the application of
the Goodenough-Kanamori-Anderson (GKA) rules for the
superexchange\cite{GKA1,GKA2,GKA3} lead to the correct assignment of magnetic
couplings in clinoclase. Therefore, all geometrical details of relevant
exchange pathways should be taken into account and considered simultaneously.
We elaborate on this problem by determining the positions of hydrogen atoms and
analyzing their role in the superexchange.

The outline of the paper is as follows. Sec.~\ref{sec:methods} summarizes
experimental and computational techniques that were applied in this study. In
Sec.~\ref{sec:sample}, we report details of the sample characterization
followed by the determination of hydrogen positions and the discussion of the
crystal structure in Sec.~\ref{sec:structure}. Sec.~\ref{sec:exp} presents
experimental magnetic properties and their brief discussion in terms of a
phenomenological model of two spin dimers with different energy scales, whereas
Sec.~\ref{sec:mag_mod} provides a microscopic insight into this model and into
residual interactions between the spin dimers. Finally, Sec.~\ref{sec:hydrogen}
clarifies the role of hydrogen atoms in the Cu--O--Cu superexchange. Our work
is concluded with a discussion and summary in Sec.~\ref{sec:disc}.

\section{Methods}
\label{sec:methods}
For our experimental studies, we used a natural sample of clinoclase from Wheal Gorland, St. Day United Mines (Cornwall, UK), which is the type locality of this rare mineral. The sample provided by the mineralogical collection of Salzburg University (Department of materials engineering and physics) features bulky dark-blue crystals of clinoclase together with smaller light-blue crystals of liroconite, Cu$_2$Al(AsO$_4$)(OH)$_4\cdot$4(H$_2$O). The crystals of clinoclase were manually separated from foreign phases and carefully analyzed by powder x-ray diffraction (XRD) and chemical analysis. Laboratory powder x-ray diffraction (XRD) data were collected using the Huber G670 Guinier camera (CuK$_{\alpha\,1}$ radiation, ImagePlate detector, $2\theta\,=\,3-100^{\circ}$ angle range). Additionally, high-resolution XRD data were collected at the ID31 beamline of the European Synchrotron Radiation Facility (ESRF, Grenoble) at a wavelength of about 0.43\,\r{A}. The chemical composition was determined by the ICP-OES method.\footnote{ICP-OES (inductively coupled plasma optical emission spectrometry) analysis was performed with the VISTA instrument from Varian.} 
The thermal stability of clinoclase was investigated by thermogravimetric analysis\footnote{Measurements are done using a Netzsch STA 449C instrument with a Al$_2$O$_3$ crucible, argon gas and a gas flow of 120~ml/min.} up to 500~$^{\circ}$C.

Magnetic susceptibility of clinoclase was measured with a Quantum Design MPMS SQUID magnetometer in the temperature range of $2-380$\,K in fields up to 5\,T. 

Electronic structure calculations were performed within density functional theory (DFT) by using the full-potential local-orbital code \textsc{fplo9.07-41}.\cite{fplo} Local density approximation (LDA)\cite{pw92} and generalized gradient approximation (GGA)\cite{pbe96} were used for the exchange-correlation potential, together with a well converged $k$-mesh of 6$\times$6$\times$6 points for the crystallographic unit cell of clinoclase and about 100 points in supercells. Hydrogen positions missing in the presently available crystallographic data\cite{ghose1965,eby1990} were obtained by structural optimizations with a threshold for residual forces of 0.002~eV/\r{A}. 

The effects of strong electronic correlations, typical for cuprates, were considered by mapping the LDA bands onto an effective tight-binding (TB) model. The transfer integrals $t_i$ of the TB-model are evaluated as nondiagonal matrix elements in the basis of Wannier functions (WFs).\cite{} These transfer integrals $t_i$ are further introduced into the half-filled one-orbital Hubbard model $\hat{H}=\hat{H}_{TB}+U_{\text{eff}}\sum_{i}\hat{n}_{i\uparrow}\hat{n}_{i\downarrow}$ that is eventually reduced to the Heisenberg model for the low-energy excitations, 
\begin{equation} 
\hat{H}=\sum_{\left\langle ij\right\rangle}J_{ij}\hat{S_{i}}\cdot\hat{S_{j}},
\end{equation}
where the summation is done over bonds $\langle ij\rangle$. For the half-filled case, which applies to clinoclase, the reduction to the Heisenberg model is well-justified in the strongly correlated limit $t_i\ll U_{\eff}$, with the effective on-site Coulomb repulsion $U_{\eff}$ exceeding $t_i$ by at least an order of magnitude (see Table~\ref{tJ}). This procedure yields AFM contributions to the exchange coupling evaluated as $J_i^{\AFM}=4t_i^2/U_{\eff}$.

Alternatively, full exchange couplings $J_i$, comprising FM and AFM contributions, can be derived from differences in total energies of various collinear spin arrangements, as evaluated in spin-polarized supercell calculations within density functional theory LSDA+$U$ formalism that includes a mean-field correction for correlation effects. An ``around mean field" (AMF) as well as a ``fully localized limit" (FLL) approximation for correcting the double counting were used.\cite{dcc} Both types supplied consistent results so that only the AMF results are presented here. The on-site Coulomb repulsion and on-site Hund's exchange of the Cu $3d$ orbitals are chosen as $U_d$\,=\,6.5$\pm$0.5\,eV and $J_d$\,=\,1\,eV, respectively, according to the parameter set used for several other cuprates.\cite{diaboleite,cuse2o5,dioptase}

In addition to periodic DFT calculations, we performed a series of cluster
calculations that pinpoint the effect of hydrogen atoms on the superexchange.
The cluster under consideration is based on the Cu$_2$O$_6$H$_5$ dimer and
embedded in a set of point charges, with two As ions bonded to the dimer
considered as total ion potentials (TIPs).\cite{tip1,tip2} The embedding was
chosen so that the intradimer hopping obtained from the cluster and periodic
LDA calculations match. The cluster calculations were done with the
\textsc{Orca 2.9} code~\footnote{F. Neese, ORCA version 2.9, MPI for
Bioinorganic Chemistry, Muhlheim a. d. Ruhr, Germany, 2011}$^,$\cite{orca} in
combination with a 6-311++G(d,p) basis set and a PBE0 hybrid
functional.\cite{pbe0a,*pbe0b} 

Quantum Monte Carlo (QMC) and simulations were performed using the codes
\textsc{loop}~\cite{loop} and \textsc{dirloop\_sse}~\cite{dirloop_sse} from the
software package \textsc{ALPS-1.3}.\cite{ALPS} Magnetic susceptibility was
simulated on finite lattices of $N$=2400 spins $S=\frac12$, using 40\,000
sweeps for thermalization and 400\,000 sweeps after thermalization.  For
simulations of magnetization isotherms, we used 4000 sweeps for
thermalization and 40\,000 sweeps after thermalization. Magnetization of the
``2+1'' model was simulated using the full diagonalization code from
\textsc{ALPS-1.3}.\cite{ALPS} 

\section{Results}
\subsection{Sample characterization}
\label{sec:sample}
Powder XRD confirmed the purity of our clinoclase sample. However, the ICP-OES analysis showed slight deviations from the ideal composition: 49.6(3)~wt.\% Cu, 18.3(1)~wt.\% As compared to 50.1(1)~wt.\% Cu, 19.7(1)~wt.\% As expected for \ccls. Additionally, trace amounts of Ca and S ($0.1-0.2$~wt.\%) were found. Other detectable elements, including transition metals, are below 0.03~wt.\%. 

\begin{figure*}[tbp]
\includegraphics[width=17.1cm]{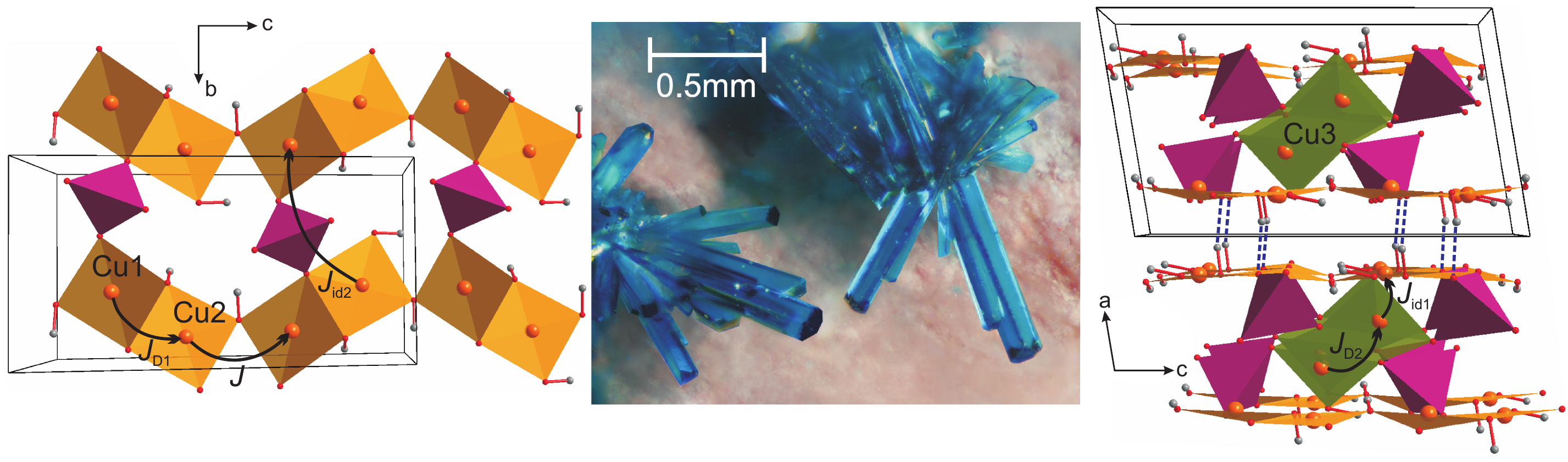}
\caption{\label{xxstr}(Color online) The crystal structure of clinoclase. The D1 (Cu1--Cu2) structural dimers are shown in orange, the D2 (Cu3--Cu3) dimers in greenish color. The left panel shows a single layer consisting of zig-zag chains of D1-dimers linked by AsO$_4$ tetrahedra (pink). In the right panel, the sandwich-like structure of clinoclase is visible with the hydrogen bonds shown as (blue) dashed lines. Arrows indicate the leading hopping pathways. The central panel shows high-quality natural clinoclase crystals from the Majuba Hill Mine, Pershing Co., Nevada, USA.}
\end{figure*}

To verify the composition of the clinoclase phase, we performed structure refinement using high-quality synchrotron data collected at room temperature. The lattice parameters we obtained for the space group $P2_1/c$ are $a=7.266$~\r A, $b=6.459$~\r A, $c=12.393$~\r A with the monoclinic angle $\beta=99.49^{\circ}$. They agree well with the existing single-crystal data,\cite{eby1990} where $a=7.257$~\r A, $b=6.457$~\r A, $c=12.378$~\r A and $\beta=99.51^{\circ}$, which we used for the magnetic modeling. Our refinement confirmed full occupation of all atomic positions in the clinoclase structure.\cite{supplement} As Cu and As are the two heaviest elements (and, therefore, strongest scatterers) in \ccls, their content in the crystalline phase is safely established by XRD. Regarding the bulk composition, the slight deficiency of Cu and As, as revealed by chemical analysis, may be attributed to trace amounts of secondary phases, such as CaCO$_3$ and CaSO$_4$ that are possible impurities in natural samples. Another plausible impurity is CuO (tenorite), which is difficult to identify by XRD, because its strongest reflections overlap with those of clinoclase. Note that none of the possible impurities reveal any conspicuous effects in the magnetic susceptibility and should not affect our experimental results reported in Sec.~\ref{sec:exp}.

Thermogravimetric analysis (TGA) identified the onset of the decomposition at about 180~$^{\circ}$C and the weight loss of 7.1\,\% upon heating to 500~$^{\circ}$C.\cite{supplement} This weight loss implies the release of 1.5 water molecules per formula unit, as expected for \ccls. The sample recovered after the heating contained a mixture of unknown phases. Their composition and crystal structures require further investigation that lies beyond the scope of the present study.

\subsection{Crystal structure}
\label{sec:structure}
Clinoclase crystallizes in the monoclinic space group $P2_1/c$ and forms a fairly complex crystal structure with three nonequivalent Cu positions (Fig.~\ref{xxstr}).\cite{eby1990} All Cu sites have a five-fold square-pyramidal coordination, where the four oxygen atoms in the basal plane form shorter bonds to Cu ($1.9-2.1$~\r A), whereas the distance to the axial oxygen atom is above 2.3~\r A. Similar to other cuprates, superexchange pathways and ensuing magnetic interactions can be described by resorting to the planar CuO$_4$ coordination, because the oxygen atom in the axial position of the pyramid does not take part in the superexchange (see Sec.~\ref{sec:mag_mod}).

The CuO$_4$ plaquettes around the Cu1 and Cu2 sites form doubly bridged dimers (D1 = Cu1--Cu2) that share corners and build zigzag chains directed along [001]. The AsO$_4$ tetrahedra connect these chains into layers in the $bc$ plane. Two layers of this type form a ``sandwich'' encompassing the dimers of Cu3 atoms (D2 = Cu3--Cu3). Although both D1 and D2 dimers are built of two CuO$_4$ plaquettes sharing a common edge, their symmetries and geometrical parameters are different. For example, the inversion symmetry in the center of D2 entails two equal Cu3--O--Cu3 angles. In contrast, D1 has no symmetry elements, hence its two Cu1--O--Cu2 bridging angles are not equal.

The connections between the ``sandwiches'' are restricted to hydrogen bonds and to long Cu--O bonds in the axial positions of the CuO$_5$ pyramids. This weak interlayer bonding is responsible for the perfect cleavage of clinoclase crystals parallel to (100), and for the low (Mohs)-hardness of 2.5--3 in this material.\cite{mohs}

Previous structure determinations\cite{ghose1965,eby1990} were based on XRD data and did not report the positions of hydrogen atoms. However, precise positions of all atoms, including hydrogen, are required for DFT band structure calculations and ensuing microscopic analysis of the electronic and magnetic structure. Therefore, we determined the hydrogen positions by relaxing the crystal structure of clinoclase. Only the hydrogen positions were optimized, whereas all other atoms were fixed to their experimental positions.\footnote{Note that we use plain LDA and GGA instead of the DFT+$U$ methods, because strong correlations in the Cu $3d$ shell should not affect the O--H bonds and the positions of hydrogen.} In the starting model, hydrogens were attached to three out of seven oxygen atoms at a typical O--H distance of 1.0~\r A. The orientation of the O--H bonds was random, although we made sure that the hydrogens are well separated from other atoms in the clinoclase structure. While there is freedom in choosing three oxygen atoms forming covalent bonds to hydrogen, only those oxygens that do not belong to the AsO$_4$ tetrahedra led to structures with low energies. When hydrogens are attached to oxygens belonging to the AsO$_4$ tetrahedra, the energy is much higher, hence such structures can be ruled out. This is in agreement with the empirical assignment of the OH groups in the experimental crystallographic study.\cite{eby1990}

The resulting hydrogen positions are listed in Table~\ref{H_opt}. Further details of the relaxation procedure and comparisons to the experiment for other Cu$^{2+}$ hydroxy-salts can be found in Refs.~\onlinecite{diaboleite,malachite}. The optimized O--H distances are close to 1.0~\r A, as expected for the covalent \mbox{O--H} bonds. Each hydrogen atom also forms one longer contact of about 1.8~\r A (hydrogen bond) to another oxygen atom. Two of these contacts provide additional bonding within the layer,\cite{supplement} whereas the hydrogen bond formed by H2 connects adjacent ``sandwiches'' (see Fig.~\ref{xxstr}). This arrangement of hydrogen bonds correlates with the positions of the hydrogens: while H1 and H3 are nearly coplanar with Cu and O atoms, H2 is notably shifted along the $a$ direction toward the neighboring ``sandwich''. Therefore, H1 and H3 lie in the plane of the D1 dimer plaquettes, whereas H2 and the respective O--H bond are in the out-of-plane position. 
 
\begin{table}[tbp]
\begin{ruledtabular}
\caption{\label{H_opt} 
Hydrogen positions obtained by LDA/GGA structure optimization. The positions of oxygen forming short O--H bonds to these hydrogen atoms are given in brackets.}
\begin{tabular}{c c c c}
atom & $x/a$ & $y/b$ & $z/c$ \\ \hline
H1 (O5) & 0.7429/0.7493 & 0.3513/0.3515 & 0.4799/0.4807 \\ 
H2 (O6) & 0.9362/0.9346 & 0.4685/0.4680 & 0.6792/0.6786 \\ 
H3 (O7) & 0.1522/0.1511 & 0.1635/0.1648 & 0.4876/0.4864 \\
\end{tabular}
\end{ruledtabular}
\end{table}

\subsection{Magnetization measurements and phenomenological fits} 
\label{sec:exp}
\begin{figure}
\includegraphics[width=8.6cm]{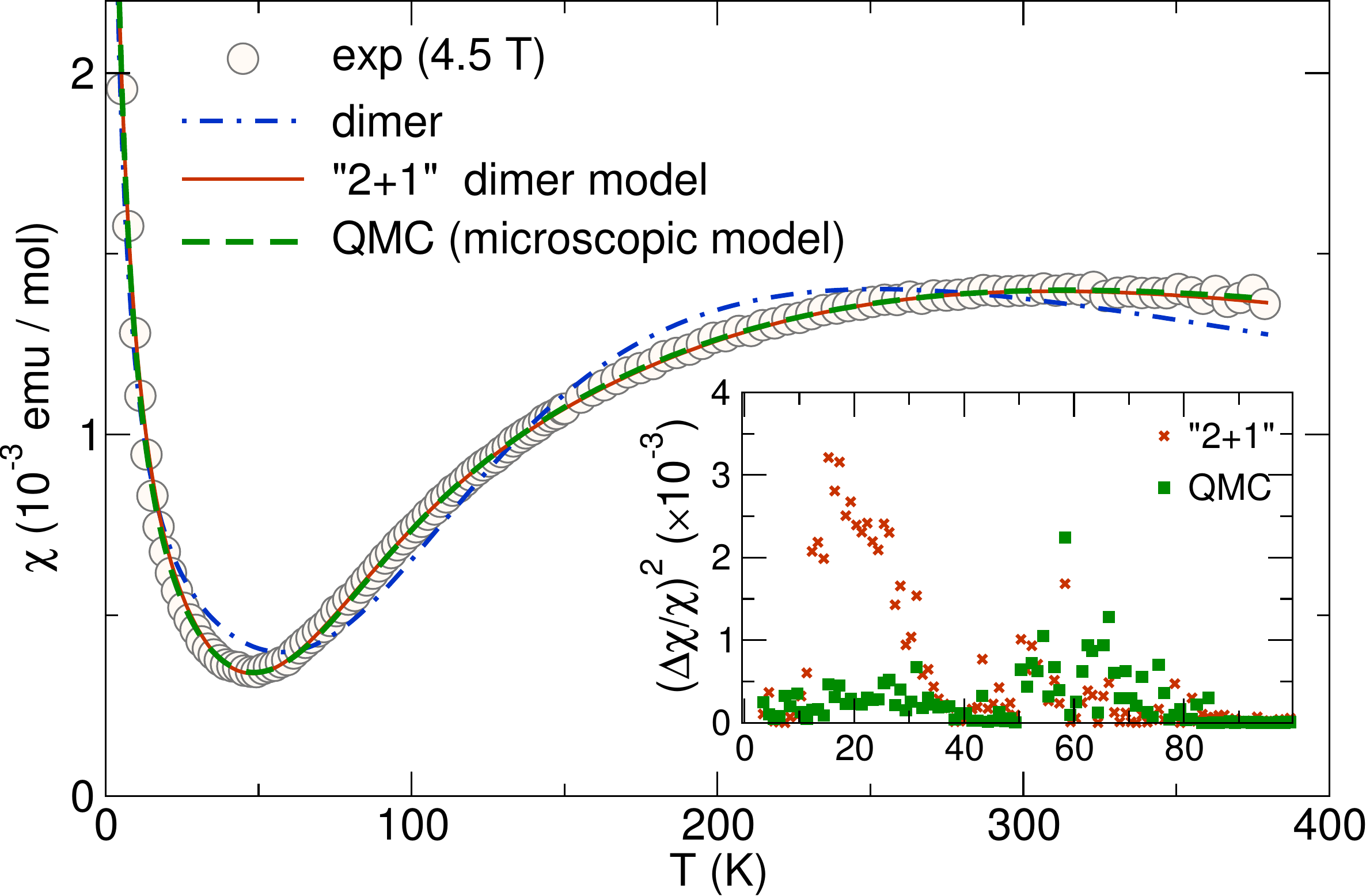}
\caption{\label{F-chi} Experimental magnetic susceptiblity (circles) of
clinoclase Cu$_3$(AsO$_4$)(OH)$_3$ and fits using a dimer model (dash-dotted
line), the phenomenological model $[$Eq.~\ref{E-H_three_dimers}$]$ of 2+1
dimers (solid line), and the microscopic model of coupled spin dimers, as shown in
Fig.~\ref{F-model} (dashed line, QMC fit). Besides the intrinsic dimer susceptibility
$[$Eq.~\ref{E-chi_dimer}$]$, we accounted for the temperature-independent
contribution and a Curie-Weiss impurity and/or defect contribution. Inset:
difference curves for the 2+1 dimers solution and the QMC fit.}
\end{figure}

The temperature dependence of the magnetic susceptibility $\chi(T)$ is shown in
Fig.~\ref{F-chi}. In quantum magnets, $\chi(T)$ typically has an asymmetric
dome-like shape, with a broad maximum indicating a gradual crossover from the
high-temperature (paramagnetic) to the low-temperature (correlated) regime. Due
to the unusually high magnetic energy scale in clinoclase, this maximum is
shifted to high temperatures ($\sim$300\,K) and is barely visible in the data
collected below and around room temperature. Unfortunately, high-temperature
measurements are not possible because the decomposition of clinoclase starts at
about 450\,K.\cite{supplement}

The upturn in $\chi(T)$ below  50\,K is a typical extrinsic feature caused by
defects and/or impurities.  It can be reasonably described by a Curie-Weiss law
with $C_{\text{imp}}\!=\!0.015$\,emu\,K\,/\,mol, corresponding to 3.2\% of
$S\!=\!\frac12$ impurities per f.\ u., and $\theta_{\text{imp}}\!=\!2.5$\,K.
After subtraction of the extrinsic contribution, we obtain vanishingly small
susceptibility below 30\,K and an activated (exponential) behavior at higher
temperatures, evidencing the gapped nature of the magnetic excitation spectrum.

The gap between the lowest-lying $S\!=\!0$ and $S\!=\!1$ states (the spin gap)
is inherent to numerous magnetic models. The simplest one is a quantum-mechanical spin dimer with a singlet ground state. Indeed, the structure of
clinoclase features well-defined structural dimers D1 and D2 that are evocative
of the spin-dimer magnetism. For a system of isolated dimers, the magnetic
susceptibility is given by the exact analytical expression:
\begin{equation} 
\label{E-chi_dimer}
\chi(T) =
\frac{Ng^2{\mu}_{\text{B}}^2}{T}\frac{1}{\left(3+\exp{[J/T]}\right)},
\end{equation} 
where $J$ is the magnetic exchange within the dimer. The fit yields
$J$\,=\,415\,K, but it does not account for the shape of the experimental
curve, as shown in Fig.~\ref{F-chi}. Moreover, the resulting value of the
$g$-factor ($g\!=\!1.48$) is unrealistically small for Cu$^{2+}$.

The failure of the isolated dimer model suggests that the magnetic couplings in
clinoclase are more intricate. Prior to the microscopic evaluation
(Sec.~\ref{sec:mag_mod}), we attempt to describe the $\chi(T)$ dependence with
a phenomenological magnetic model. According to Sec.~\ref{sec:structure}, the
monoclinic unit cell of clinoclase contains three structural dimers formed by
edge-sharing CuO$_4$ plaquettes (Fig.~\ref{xxstr}). Two of these dimers are of
the D1 type (Cu1--Cu2), whereas the third dimer is D2 (Cu3--Cu3). Naturally,
the magnetic couplings within D1 are different from those in D2. This brings us
to a tentative model of 2+1 dimers:
\begin{subequations}
\label{E-H_three_dimers}
\begin{eqnarray}
H = 2H_{\Done} + H_{\Dtwo} \\
H_{\Done} = \frac12\sum_{\substack{i\in\text{Cu1}\\j\in\text{Cu2}}}J_{\Done}(\mathbf{S}_i\cdot\mathbf{S}_j)\\
H_{\Dtwo} = \sum_{\langle{}i,j\rangle\in\text{Cu3}}J_{\Dtwo}(\mathbf{S}_i\cdot\mathbf{S}_j),
\end{eqnarray}
\end{subequations}
where $J_{\Done}$ and $J_{\Dtwo}$ denote the magnetic couplings in the
respective dimers. For each dimer, the magnetic susceptibility is given by
Eq.~(\ref{E-chi_dimer}).

Two scenarios are possible: either $J_{\Done}>J_{\Dtwo}$, or the other way
around. Fitting to the experimental curve readily shows that the
$J_{\Dtwo}\!>\!J_{\Done}$ solutions do not reproduce the experimental behavior.
In contrast, the model with $J_{\Done}\!>\!J_{\Dtwo}$ yields an excellent fit
with $J_{\Done}$\,=\,703.5\,K $J_{\Dtwo}$\,=\,289.3\,K, and $g\!=\!1.86$ (full
line in Fig.~\ref{F-chi}). Therefore, the smaller gap of $\sim$290\,K comes
from the Cu3--Cu3 dimers (D2), while the Cu1--Cu2 dimers (D1) give rise to the
larger gap of $\sim$700\,K.

\subsection{Microscopic magnetic model}
\label{sec:mag_mod}
\begin{figure}[tbp]
\includegraphics[width=8.6cm]{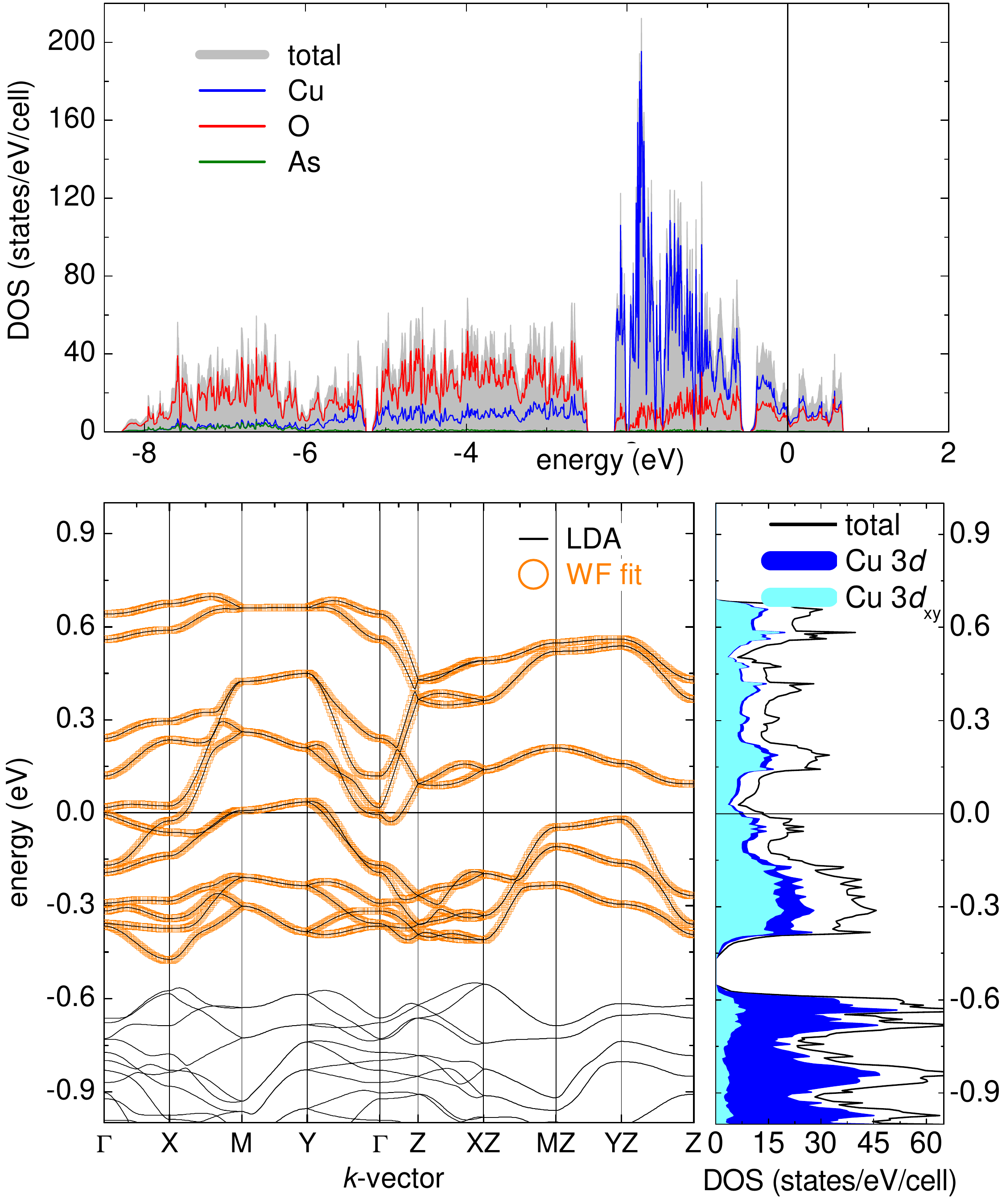}
\caption{\label{bands}
(Color online) The LDA density of states (DOS) and the band structure of clinoclase
\ccls.  The top panel shows the contributions of the Cu(3$d$), O(2$p$) and As
states to the total DOS. The Fermi level is at zero energy.  In the bottom left
panel, the LDA-bands around the Fermi level are displayed and compared with
bands derived from a fit using an effective one-band tight-binding model based
on Cu-centered Wannier functions (WFs) projected on local
Cu(3$d_{xy}$)-orbitals. The $k$-points are defined as follows: $\Gamma=(000)$, X$=(\frac{\pi}{a}00)$,
Y$=(0\frac{\pi}{b}0)$, Z$=(00\frac{\pi}{c})$, XZ$=(\frac{\pi}{a}0\frac{\pi}{c}), $MZ$=(\frac{\pi}{a}\frac{\pi}{b}\frac{\pi}{c})$, YZ$=(0\frac{\pi}{b}\frac{\pi}{c})$. The bottom right panel shows that the partial Cu(3$d$)-DOS at the Fermi level is basically of Cu(3$d_{xy}$) character,
justifying our construction of the WFs. }
\end{figure}

Now, we compare the above phenomenological model with the microscopic results based on DFT. 
In a first step, LDA calculations are performed. The width of the whole valence
band block of about 9\,eV is typical for cuprates (see Fig.~\ref{bands}). The spurious metallicity of the energy spectrum is a well-known shortcoming of LDA due to the underestimated electronic correlation in the Cu $3d$ shell.\footnote{LSDA+$U$ calculations arrive at the band gap of about 2.6~eV in reasonable agreement with the blue color of clinoclase.} Nevertheless, the LDA bands around the Fermi level (the energy range from $-0.5$ to 0.7\,eV) are sufficient to describe low-energy magnetic excitations, provided that a suitable correlation part is added to the model Hamiltonian. The relevant bands
are essentially of the Cu $d_{xy}$ character, with sizable contributions from
O 2$p$ orbitals. The orbital symmetry is defined with respect to the local
coordinate system on each CuO$_4$-plaquette, where the Cu--Cu bond of the dimer
is chosen as the local $x$-axis, and the $z$-axis is orthogonal to the plaquette plane. Note that this setting is different from the standard one, where $x$ and $y$ axes are directed along the Cu--O bonds, so that the highest crystal-field level has the $d_{x^2-y^2}$ symmetry.

The leading hopping parameters $t_i$ and corresponding AFM exchanges $J^{\text{AFM}}_i=4t_i^2/U_{\eff}$ are listed in Table~\ref{tJ}. The results of the model analysis are supported by the evaluation of full exchange integrals $J_i$ using total energies of collinear spin configurations calculated with LSDA+$U$. These two approaches are complementary. The model analysis provides information on all exchange couplings in the system and guides the LSDA+$U$ calculations that are restricted to only a handful of leading interactions.

Our model analysis based on the hopping parameters $t_i$ identifies five leading AFM exchanges that exceed 100~K (see $J_i^{\AFM}$ in Table~\ref{tJ}). The perfect agreement between the LDA bands at the Fermi level and those calculated with the Cu-centered Wannier functions (Fig.~\ref{bands}) confirms that the relevant superexchange pathways in clinoclase can be well described in terms of the CuO$_4$ plaquettes. Despite the short Cu--Cu distance of only 3.3~\r A between the Cu atoms in two contiguous ``sandwiches'' along the $a$ direction, the respective hopping is negligibly small (below 20~meV) because the magnetic orbitals lie in the $bc$ plane and do not overlap. Likewise, two outer layers of each ``sandwich'' are coupled only via the Cu3 spins and lack any direct interaction.

\begin{table}[tbp]
\begin{ruledtabular}
\caption{\label{tJ}
Leading exchange couplings in clinoclase \ccls: Cu-Cu distances $d_{\text{Cu-Cu}}$ (in~\r{A}), bridging angles $\varphi_{\text{Cu--O--Cu}}$ (in~deg), hopping integrals $t_i$ (in~meV), antiferromagnetic contributions $J^{\AFM}_i=4t^2/U_{\text{eff}}$ (in~K) with $U_{\text{eff}}$\,=\,4.0\,eV. The $J_i$ (in~K) are calulated with LSDA+$U$ using $U_d=6.5\pm0.5$eV, $J_d=1$eV.}
\begin{tabular}{l c c r r r}
 &  $d_{\text{Cu--Cu}}$ & $\varphi_{\text{Cu--O--Cu}}$ & $t_i$ & $J^{\AFM}_i$ & $J_i$  \\ \hline
$J_{\Done}$     & 2.98 & 93.6/99.9 & $-115$ & 153 & $-4\pm8$   \\ 
$J_{\Dtwo}$     & 3.13 & 101.9   &  191   & 423 & $302\pm53$ \\
$J_{id1}$       & 3.38 & 124.6     & $-117$ & 159 & $161\pm25$ \\
$J$             & 3.66 & 149.3     &  276   & 884 & $693\pm99$ \\
$J_{id2}$       & 5.52 &  --       & $-106$ & 130 & $159\pm31$ \\
\end{tabular}
\end{ruledtabular}
\end{table}
 
The comparison between $J_i^{\AFM}$ and $J_i$ in Table~\ref{tJ} shows that four out of five leading interactions are indeed AFM, with only small FM contributions. However, the coupling within D1 is nearly canceled because of comparable FM and AFM terms. The large FM contribution to $J_{\Done}$ ($J_{\Done}^{\FM}=J_{\Done}-J_{\Done}^{\AFM}\simeq -160$~K) is indeed expected for the coupling geometry with bridging angles close to $90^{\circ}$. 

Our microscopic model is consistent with the phenomenological analysis that suggested spin dimers with two different energy scales (Sec.~\ref{sec:exp}). The coupling on D2 is $J_{\Dtwo}=302$~K very close to 300~K found experimentally. However, the magnetic dimer with the coupling of about 700~K has to be re-assigned. The coupling within D1 is in fact very weak, so that the spin dimer with $J\simeq 700$~K is formed not on the Cu1--Cu2 structural dimer D1, but between the respective dimers, where the CuO$_4$ plaquettes share a common corner instead of sharing a common edge. This effect can be well understood in terms of the GKA rules for the superexchange, because the Cu--O--Cu angle for $J$ is nearly 150$^{\circ}$ compared to only $93-100^{\circ}$ for $J_{\Done}$. However, the GKA rules do not account for the fact that $J_{\Dtwo}\simeq 302$~K exceeds $J_{id1}\simeq 159$~K, even though the bridging angle for $J_{\Dtwo}$ is notably smaller. This unusual behavior is further discussed in the next sections.

The re-assignment of the magnetic dimer has no effect on the fit of the magnetic susceptibility presented in Sec.~\ref{sec:exp} since it is independent of the position of the dimers in the crystal structure. The interdimer couplings $J_{id1}$ and $J_{id2}$ are non-frustrated and can be taken into account by QMC. The resulting fit shown in Fig.~\ref{F-chi} is only slightly better than the fit with the phenomenological ``2+1'' model. We find $J=706.8$~K, $J_{\Dtwo}=318.1$~K and $g=1.893$ in good agreement with our previous results. The coupling between the magnetic dimers were chosen as \hbox{$J_{id1}$\,=\,$J_{id2}$\,=\,0.125$\,J$} to yield best agreement with the experimental curve.

\begin{figure}[tbp]
\includegraphics[width=8.6cm]{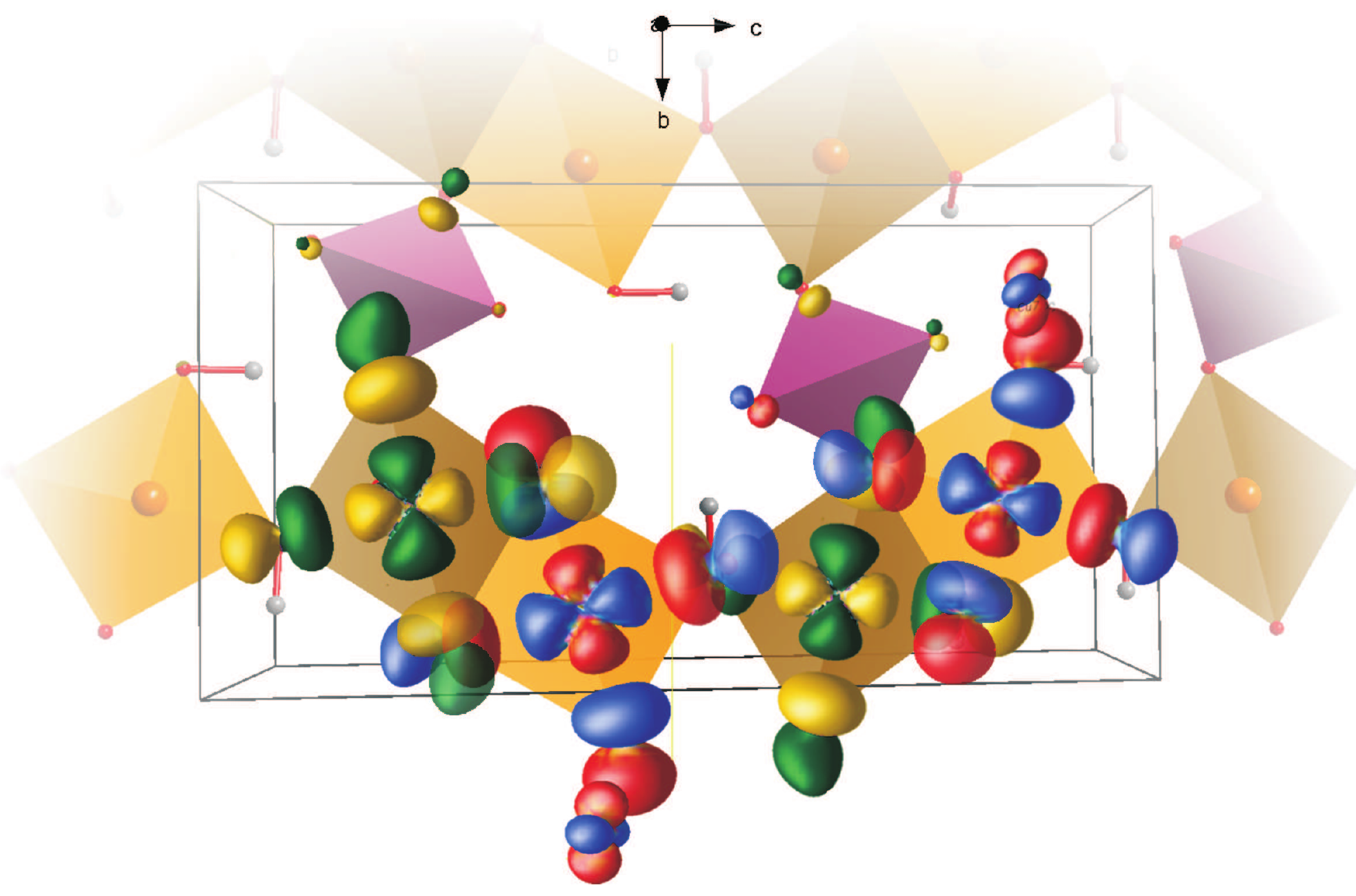}
\caption{\label{WFrot}
(Color online) Wannier functions (WFs) on the Cu1 (yellow-green) and Cu2 sites (red-blue). The net overlap at the two bridging oxygen within the D1 dimers is significantly smaller than the overlap at the oxygen bridging two dimers. The Cu1-WF exhibits a considerable distortion towards the AsO$_4$-tetrahedra responsible for the large $J_{id2}$ coupling. }
\end{figure}

\begin{figure}[tbp]
\includegraphics[width=8.6cm]{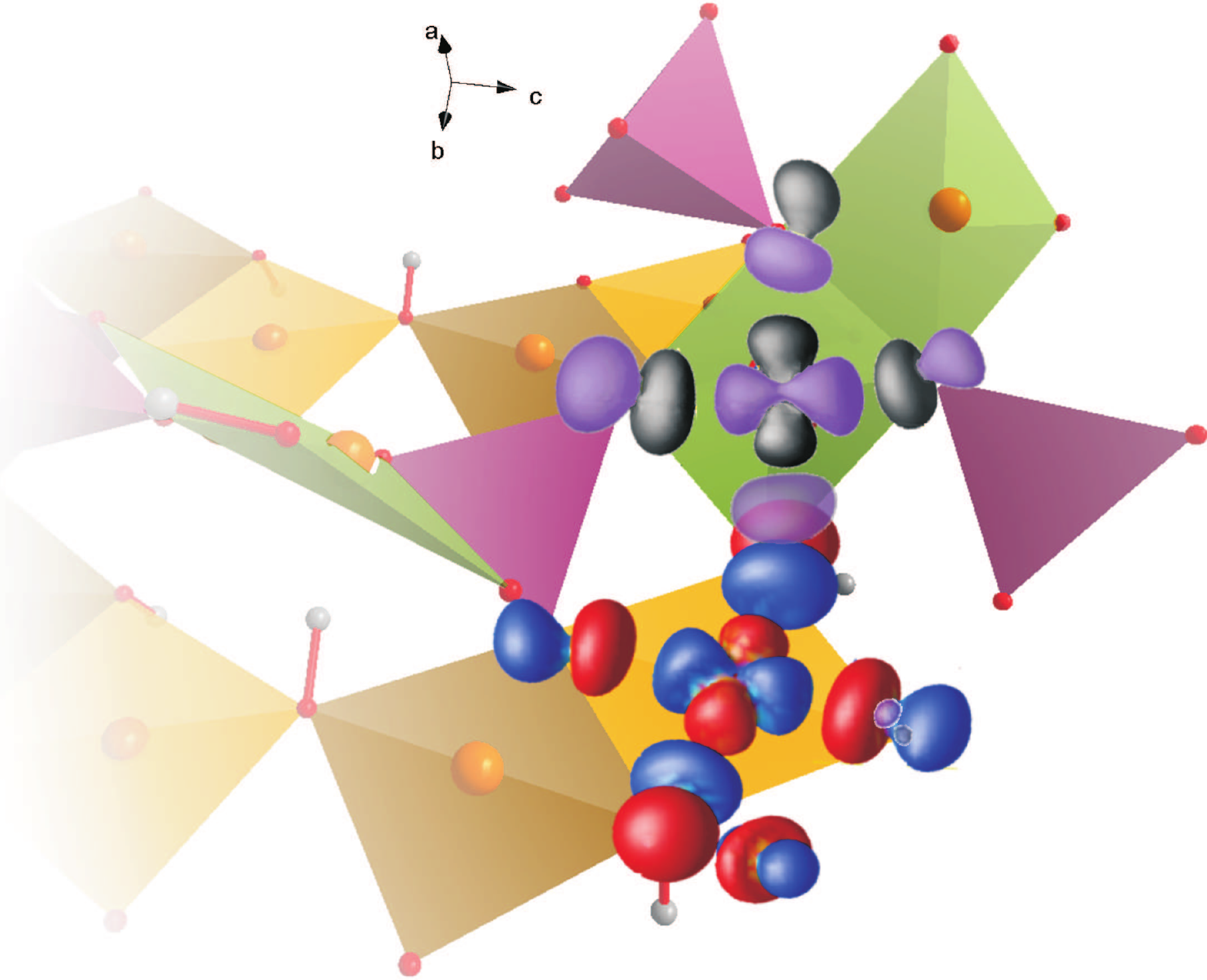}
\caption{\label{wf1}
(Color online) Wannier functions (WFs) on the Cu2 (red-blue) and Cu3 (grey-violet). The overlap between Cu2 and Cu3 is hampered by the non-planar arrangement of the CuO$_4$ plaquettes. The Cu3-WF features distortions of the O(2$p$) contributions caused by the AsO$_4$ tetrahedra. }
\end{figure}

\begin{figure}[tbp]
\includegraphics[width=8.6cm]{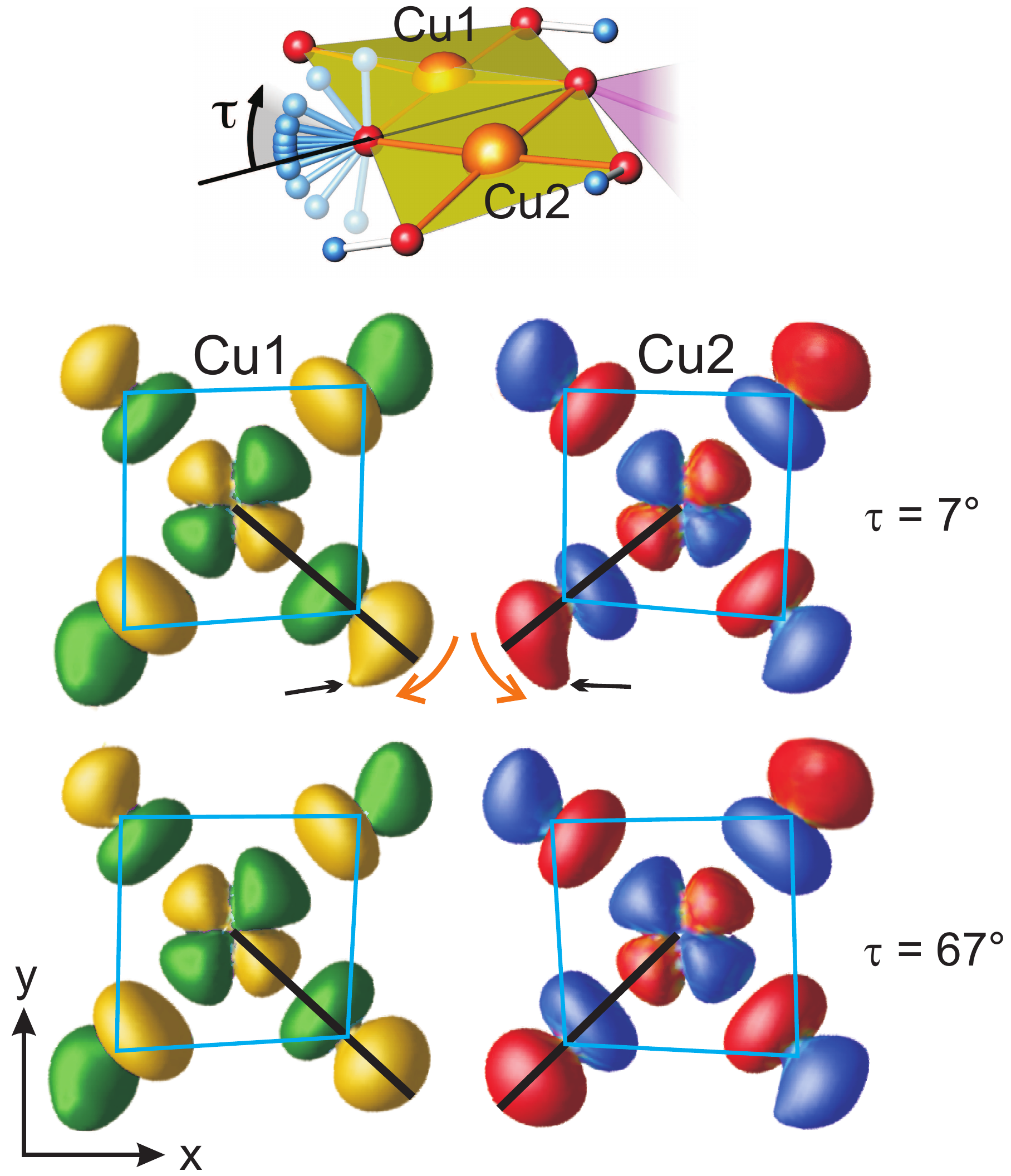}
\caption{\label{wf2}
(Color online) The effect of the out-of-plane rotation of hydrogen on the Wannier functions (WFs) localized at the Cu1 and Cu2 sites. The top figure shows the D1 structural dimer with H, bonded to the bridging oxygen, being rotated out of the dimer plane by an angle $\tau$. The central and bottom figures show the WFs at the two Cu sites for $\tau=7^{\circ}$ and $\tau=67^{\circ}$, respectively. The CuO$_4$ plaquettes are indicated in light blue, black lines connect the central Cu and the O bonded to H. With increasing $\tau$, the O(2$p$) orbital rotates (indicated by orange arrows) about a local $z$-axis, which is visible with respect to the black line. Black arrows point at the H(1$s$) contribution to the WFs. At $\tau=67^{\circ}$ that corresponds to the computationally relaxed structure, no such contributions are visible anymore.}
\end{figure}

\subsection{Role of hydrogen}
\label{sec:hydrogen}
One important difference between the coupling pathways for $J_{\Done}$, $J_{\Dtwo}$, and $J_{id1}$ pertains to the positions of hydrogen atoms. While the bridging oxygen atoms of D2 belong to the AsO$_4$ tetrahedra and have only weak contacts to hydrogen, the bridging atom for $J_{id1}$, as well as one of the bridging atoms of D1, form covalent O--H bonds. 
For studying the role of the O--H bonds in more detail, we will focus on the D1 dimer where, according to the very small exchange coupling and the planar Cu$_2$O$_6$ geometry, we expect interesting effects, as, e.g., a change from FM to AFM coupling.

The effect of the out-of-plane angle $\tau$ of the O--H bond on the intradimer coupling was studied by Ruiz \textit{et al}.\cite{ruiz97_2} for small organic ligands. They found that a large $\tau$ (out-of-plane position of hydrogen) favors FM coupling. We attempted to verify this effect for the D1-dimer in clinoclase. In a first step, the $t_{\Done}$ hopping parameters are calculated as a function of $\tau$ in the periodic model. The out-of-plane rotation of H up to 67$^{\circ}$, which corresponds to the optimized crystal structure of clinoclase, reduces $t_{\Done}$ by about 40\% and, thus, the AFM contribution to $J_{\Done}$ by about
60\% (see Supplemental material\cite{supplement}). Furthermore, $t$ is slightly reduced and the inter-sandwich hoppings decrease by about 50\%. All other hoppings
are more or less independent of $\tau$.  

For the calculation of $J_{\Done}$ as a function of $\tau$, we used the Cu$_2$O$_6$H$_5$ cluster model, embedded in TIPs and point charges, which allows to investigate the intradimer coupling exclusively. Additionally, the cluster enables us to vary the bridging angles without changing the whole set of additional structural parameters, as this would be the case for a periodic
model.\cite{cluster} 

\begin{figure}[tbp]
\includegraphics[width=8.6cm]{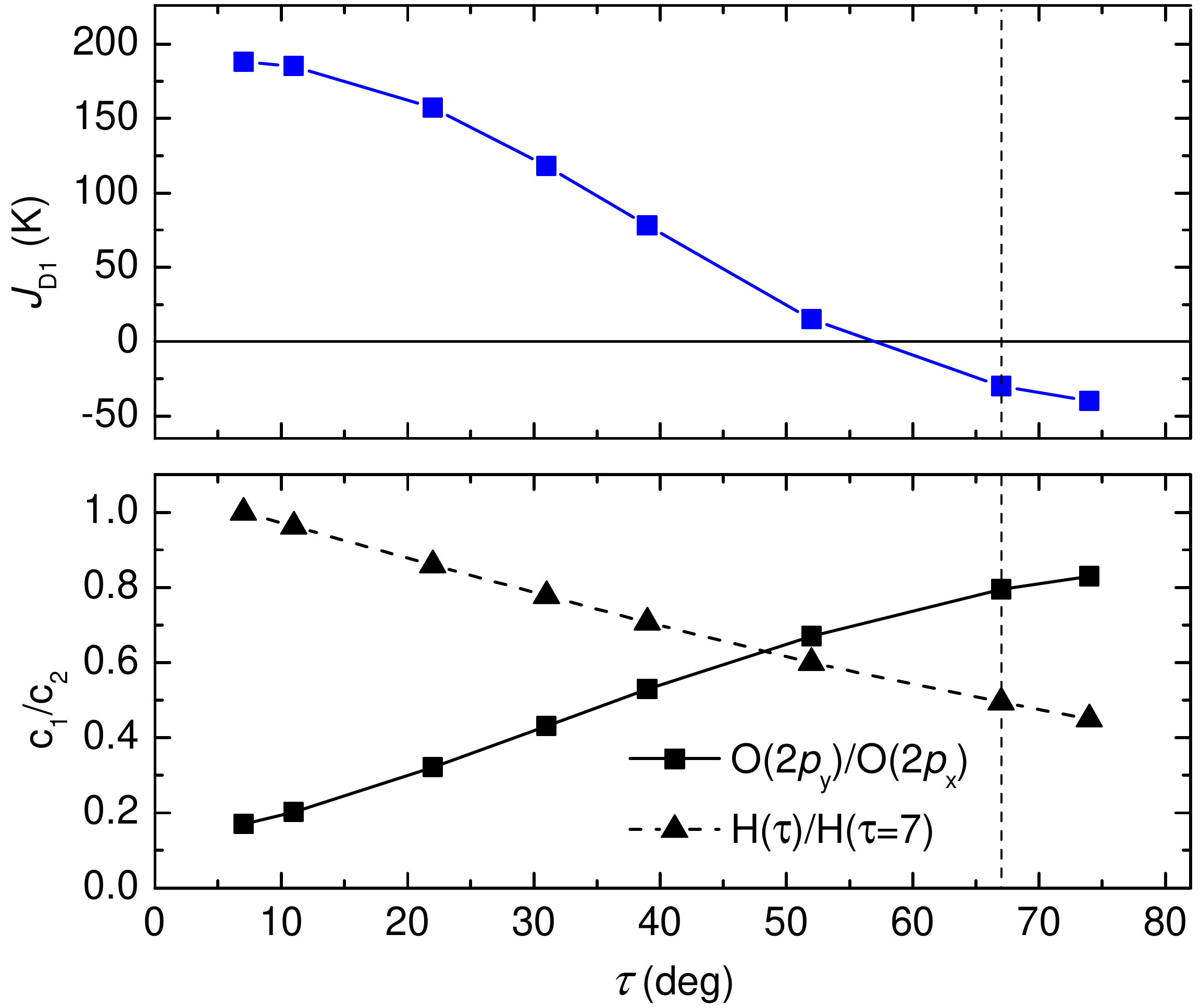}
\caption{\label{rotH_J}
(Color online) The upper panel shows $J_{\Done}$ as a function of the out-of-plane angle of hydrogen, $\tau$. In the lower panel, the ratio O(2$p_y$)/O(2$p_x$) of the WF-contributions of the bridging oxygen bonded to H are displayed. The orbital character is denoted with respect to the coordinate system shown in Fig.~\ref{wf2}. H($\tau$)/H($\tau=7$) shows the H(1$s$) contribution normalized to the value at $\tau=7^{\circ}$. The vertical dashed line indicates the computationally relaxed out-of-plane angle of hydrogen.}
\end{figure}

The results of the cluster calculations (Fig.~\ref{rotH_J}) nicely show the transition from AFM to FM coupling upon an increase in $\tau$. This effect is driven by the reduced hopping, because the FM contribution $J^{\FM}_{\Done}=J_{\Done}-J^{\AFM}_{\Done}$ is weakly dependent on $\tau$ and hovers around $-150$~K. The absolute size of $J_{\Done}$ obtained in the cluster calculations with the PBE0 functional is somewhat larger than the LSDA+$U$ estimates. This is in fact little surprising, because hybrid functionals, such as PBE0, tend to overestimate the exchange couplings.\cite{tip2}

The decrease in $t_{\Done}$ can be traced back to the increasing contribution of the bridging oxygen O(2$p_y$) to the WFs of the Cu1 and Cu2 sites (Fig.~\ref{rotH_J}), while the O(2$p_x$) contribution remains constant. At small $\tau$, H(1$s$) strongly interacts with O(2$p_y$) and thus shifts its orbital energy downwards, which in turn reduces the interaction between this oxygen orbital and Cu(3$d_{xy}$) orbitals. The H(1$s$) contribution itself decreases with increasing $\tau$. This is also visible in the WF-picture (Fig.~\ref{wf2}) as the rotation of the contribution of bridging oxygen atoms. As the $\tau$ increases, the O(2$p$) orbital turns into the direction perpendicular to the Cu--Cu axis of D1, hence the overlap of the WF's of Cu1 and Cu2 is reduced.

\begin{figure}
\includegraphics[width=8.6cm]{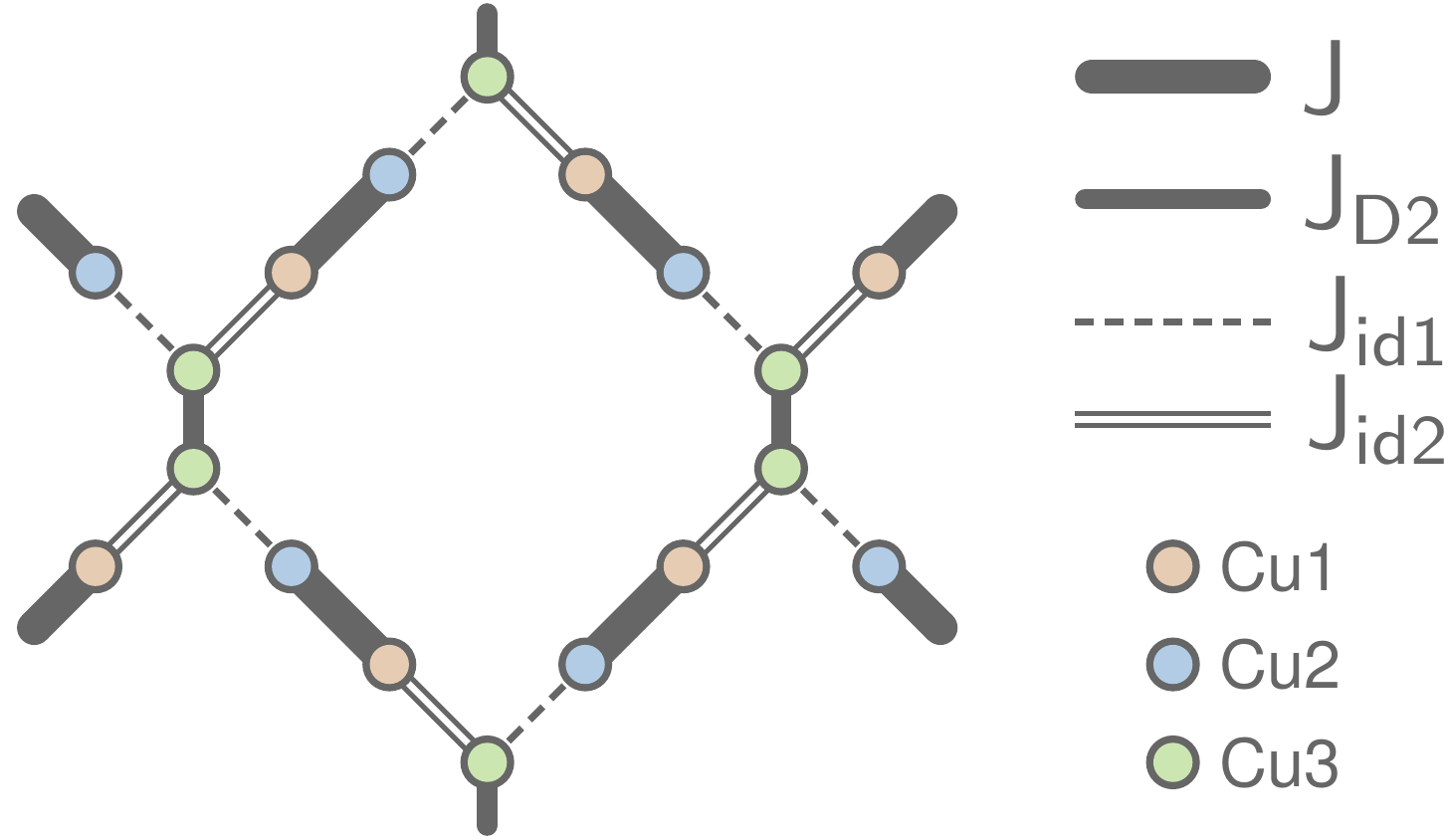}
\caption{\label{F-model} Microscopic magnetic model of clinoclase \ccls. Note
that each structural ``sandwich'' depicted in Fig.~\ref{xxstr} comprises three
interpenetrating lattices of this type that are decoupled from each other. For
the notation of magnetic couplings, see Table~\ref{tJ}.}
\end{figure}
\begin{figure}
\includegraphics[width=8.6cm]{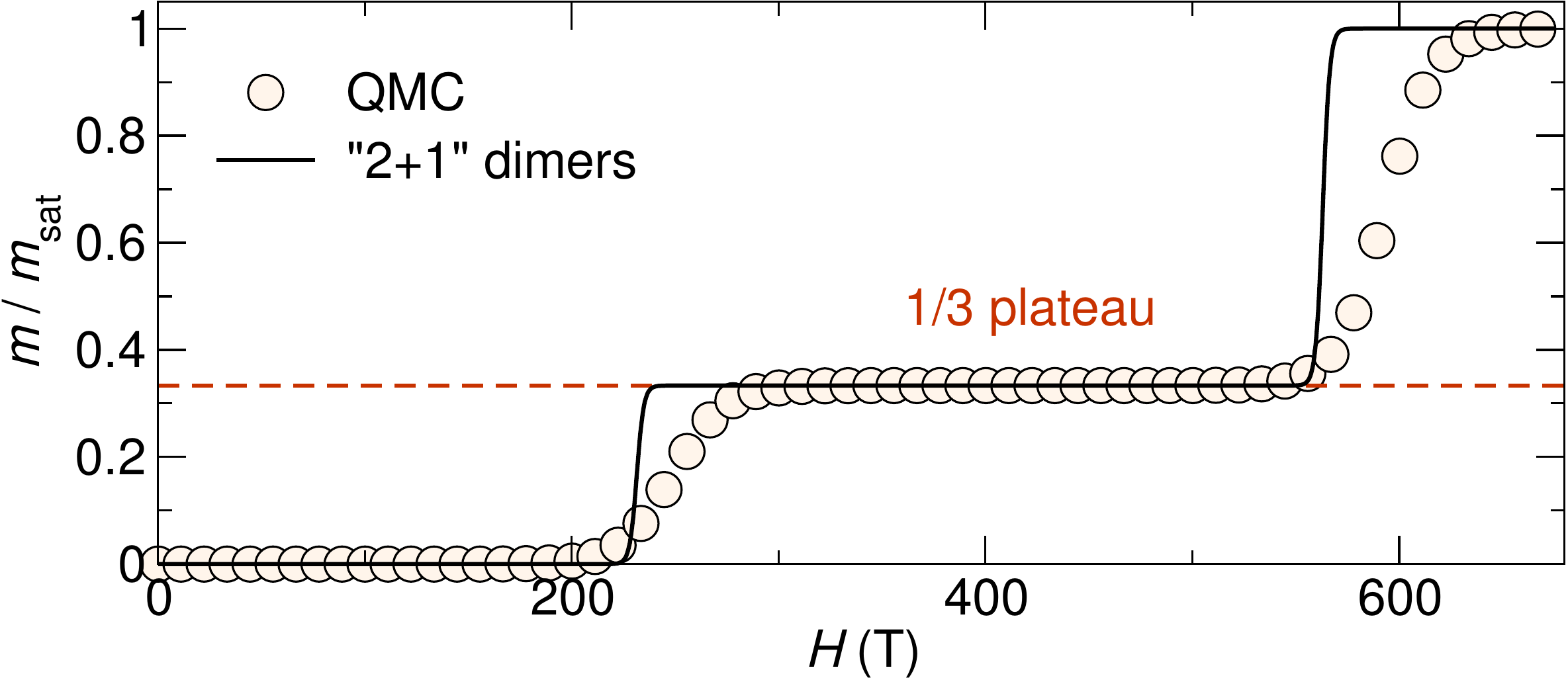}
\caption{\label{F-mh} Magnetization isotherm of clinoclase \ccls\ simulated
using exact diagonalization of the ``2+1'' dimers model and QMC for the microscopic magnetic model with
$J\!:\!J_{\Dtwo}\!:\!J_{id1}\!:\!J_{id2}$\,=\,1\,:\,0.45\,:\,0.125\,:\,0.125.
Magnetic field is scaled adopting $J$ and $g$ values from the
$\chi(T)$ fits (Fig.~\ref{F-chi}). $m_{\text{sat}}$ is the saturation magnetization. Note two wide
plateaus at $m/m_{\text{sat}}\!=\!0$ and $m_{\text{sat}}\!=\!1/3$.
}
\end{figure}

\section{Discussion}
\label{sec:disc}
The magnetism of clinoclase is well described by a model of two nonequivalent spin dimers. While this model is easily derived from the crystal structure of the mineral (Sec.~\ref{sec:exp}), the arrangement of magnetic dimers cannot be established on purely empirical grounds. Two shortest Cu--Cu distances are formed within the structural dimers D1 and D2. Assuming that the magnetic coupling is more efficient at short distances, one would identify these structural dimers as magnetic dimers. This assessment is correct for D2, yet it fails for D1, where the strong interaction $J$ forming the magnetic dimer is found between the structural D1 dimers. The GKA rules could provide a more plausible scenario, because they account for the fact that short Cu--Cu distances entail Cu--O--Cu angles close to $90^{\circ}$ that are unfavorable for an AFM coupling. The reference to the GKA rules readily explains why the dimer is formed by $J$ (bridging angle of 149.3$^{\circ}$) and not by $J_{\Done}$ (bridging angles below 100$^{\circ}$). However, a consistent application of these rules would also relegate $J_{\Dtwo}$ (101.9$^{\circ}$) that should be well below $J_{id1}$ with a much larger Cu--O--Cu angle of 124.6$^{\circ}$.

In clinoclase, neither Cu--Cu distances nor Cu--O--Cu angles fully elucidate the microscopic scenario. To explain why $J_{\Dtwo}$ exceeds $J_{id1}$, details of their superexchange pathways should be analyzed and compared. Apart from the hydrogen atoms considered in Sec.~\ref{sec:hydrogen}, we find two main differences between these couplings: i) the number of bridging oxygen atoms, which is two for $J_{\Dtwo}$ and one for $J_{id1}$; ii) the mutual arrangement of the CuO$_4$ plaquettes that are coplanar for $J_{\Dtwo}$ and strongly twisted for $J_{id1}$. Dividing the overall Cu--Cu hopping $t_{\Dtwo}$ by two, we obtain $t_{\Dtwo}^{\eff}=95$~meV that reflects the transfer via a single Cu--O--Cu bridge (as in $J_{id1}$). This hopping is slightly below $|t_{id1}|=117$~meV, but their difference is much smaller than expected for the bridging angles of 101.9$^{\circ}$ and 124.6$^{\circ}$, respectively. For example, the model calculation from Ref.~\onlinecite{braden} suggests that $t_{id1}$ should be at least twice larger than $t_{\Dtwo}$. 

There are different scenarios explaining the large AFM coupling of $J_{\Dtwo}$: It may originate from the combined effect of the indirect Cu--O--Cu and direct Cu--Cu hoppings within the Cu$_2$O$_6$ structural dimer. While the Cu--O--Cu processes should be solely determined by the bridging angle, the direct hopping requires the coplanar arrangement of the CuO$_4$ plaquettes, which is the case for $J_{\Dtwo}$ only. This explanation is in line with the robust AFM coupling observed in many other spin-dimer compounds, such as TlCuCl$_3$ (Ref.~\onlinecite{[{For example: }][{}]oosawa2002}) and SrCu$_2$(BO$_3)_2$ (Ref.~\onlinecite{miyahara2003,*takigawa2010}), despite their low bridging angles of $96-98^{\circ}$. Same arguments could be applied to D1, where one of the bridging angles is as large as 100$^{\circ}$ and indeed leads to a sizable transfer $t_{\Done}=-115$~meV. However, the out-of-plane O--H bond (see Sec.~\ref{sec:hydrogen}) has strong impact on the superexchange and is responsible for the very weak coupling. This effect of side groups may also play an important role for D2, yet in a different manner. Here, the bridging oxygen atoms belong to the AsO$_4$ tetrahedra that could amplify the AFM superexchange, similar to GeO$_4$ tetrahedra in CuGeO$_3$.\cite{geertsma1996} Note that the structural dimers in $\alpha$-Cu$_2$As$_2$O$_7$ (bridging angle of 101.7$^{\circ}$) are very similar to D2 and also feature a strong AFM coupling.\cite{arango2011,janson2011}

Another interesting feature of clinoclase is the sizable interdimer coupling $J_{id2}$. In contrast to all magnetic couplings discussed so far, $J_{id2}$ does not involve a direct connection between the CuO$_4$ plaquettes and occurs via the bridging AsO$_4$ tetrahedron. The efficiency of this superexchange pathway can be explained by the coplanar and, moreover, well aligned arrangement of the CuO$_4$ plaquettes. Their positions are such that two Cu--O$\cdots$O--Cu contacts are formed. One of these contacts goes along the edge of the AsO$_4$ tetrahedron, while the second contact does not involve any bonds or polyhedra. Nevertheless, its short O$\cdots$O distance of $2.8-3.0$~\r A is likewise beneficial for the superexchange. The resulting coupling $J_{id2}\simeq 140$~K is comparable to the typical interaction via double PO$_4$ and AsO$_4$ bridges in  Sr$_2$Cu(PO$_4)_2$,\cite{johannes2006} K$_2$CuP$_2$O$_7$,\cite{nath2008} and Cu$_2$As$_2$O$_7$.\cite{arango2011}

Regarding the spin model of clinoclase, the ``strong'' ($J$) and ``weak'' ($J_{\Dtwo}$) magnetic dimers are joined into a planar structure (Fig.~\ref{F-model}) by non-frustrated interdimer couplings $J_{id1}$ and $J_{id2}$. Three interpenetrating planes of this type together form one structural ``sandwich'' and remain nearly decoupled. Topologically, each of the three lattices represents a diluted square lattice of magnetic dimers, depicted schematically in Fig.~\ref{F-model}.

This spin lattice only marginally differs from a simple superposition of two
nonequivalent dimers. The phenomenological ``2+1'' dimer model and the 2D spin
lattice provide nearly indistinguishable fits of the magnetic susceptibility
(Fig.~\ref{F-chi}). Field dependence of the magnetization, as calculated by
QMC, is very close to the intuitive picture of isolated spin dimers
(Fig.~\ref{F-mh}). The wide plateau at $M=\frac13$ is due to the saturation of
the ``weak'' dimers ($J_{\Dtwo}$), while the ``strong'' ($J$) dimers remain in
the singlet state. Therefore, $J_{id1}$ and $J_{id2}$ have little effect on the magnetic susceptibility of clinoclase. 

The effect of these weak couplings is visible by comparing the magnetization curves simulated for isolated dimers and for the dimers coupled by $J_{id1}$ and $J_{id2}$, as shown in Fig.~\ref{F-model}. In the model augmented by $J_{id1}$ and $J_{id2}$, the transitions preceding and following the $\frac13$-plateau are broadened compared to the ``2+1'' dimer model (Fig.~\ref{F-mh}). This broadening is caused by the interdimer couplings $J_{id1}$ and $J_{id2}$ that
give rise to a dispersion of magnetic excitations, and mediate magnon-magnon
interactions underlying the peculiar effect of spontaneous magnon
decay.\cite{zhitomirsky2013} 

Clinoclase belongs to the family of gapped quantum magnets with nonequivalent
spin dimers. In contrast to other systems of this type, different spin dimers
are inherent to the crystal structure of this mineral. They are not formed due
to a symmetry reduction upon a low-temperature phase transition that keeps
similar magnetic interactions and, therefore, similar spin gaps in all dimers,
as in BaCuSi$_2$O$_6$
(Ref.~\onlinecite{sheptyakov2012,*ruegg2007,*kraemer2007}) and
NH$_4$CuCl$_3$.\cite{ruegg2004} Clinoclase can be rather compared to the
ambient-pressure modification of (VO)$_2$P$_2$O$_7$, where two distinct spin
gaps of 32~K and 65~K define two different energy scales of the
system.\cite{yamauchi1999,*kikuchi1999,kuhlmann2000,johnston2001} Systems of
this type may show interesting high-field behavior, because each group of
dimers ($J$ and $J_{\Dtwo}$) has independent low-temperature transitions
related to the Bose-Einstein condensation (BEC) of magnons. The BEC transition
takes place in the local field determined by the second group of dimers, either
unpolarized $J$ dimers for the BEC transition in $J_{\Dtwo}$, or polarized
$J_{\Dtwo}$ dimers for the BEC transition in $J$. Unfortunately, the critical
fields of clinoclase (Fig.~\ref{F-mh}) are too high to observe such effects
using present-day high-field facilities. Nevertheless, the search for similar
systems with structurally different spin dimers should be an interesting avenue
to explore the high-field physics of quantum magnets.

\section{Summary}
\label{sec:summ}
In summary, we performed a joint theoretical and experimental study of the
magnetic behavior of the mineral clinoclase. Using density functional theory
calculations, we evaluate the microscopic model for this compound and identify two types of spin dimers with the couplings of $J\simeq 700$~K and $J_{\Dtwo}\simeq 300$~K. Intuitively, one is tempted to ascribe them to two types of structural Cu$_2$O$_6$ dimers in clinoclase. In fact, only $J_{\Dtwo}$ pertains to the structural dimer D2, while the strong coupling
$J$ occurs between two corner-sharing D1 dimers. Additional couplings between the magnetic dimers reach 150~K, but play a minor role in the magnetic behavior. Simulations for the DFT-based microscopic magnetic model yield excellent agreement with
the experimental data.

The magnetic couplings in clinoclase are not solely determined by the Cu--O--Cu
angles. The AsO$_4$ side groups and the hydrogen atoms also play an important role by enhancing or suppressing antiferromagnetic contributions to short-range couplings $J_{\Done}$, $J_{\Dtwo}$, and $J_{id1}$. Since no hydrogen positions were available from the experiment, we determined them by optimizing the crystal structure within DFT. We have demonstrated that the magnetic coupling within D1 is strongly affected by the hydrogen atom attached to one of the bridging oxygens. The out-of-plane H position is responsible for the almost canceled exchange coupling $J_{\Done}$. Our findings put forward details of the crystal structure, including inconspicuous and typically overlooked effects like hydrogen positions, as an important and even decisive factor in the magnetic superexchange.

\acknowledgments
We are grateful to Gudrun Auffermann for her kind help with the chemical
analysis. We also acknowledge the experimental support by Yurii Prots and Horst
Borrmann (laboratory XRD), Yves Watier (ID31), and the provision of the ID31
beamtime by ESRF. We would like to thank the Department of Materials
Engineering and Physics of the Salzburg University for providing the
high-quality natural sample of clinoclase from their mineralogical collection
(inventory number 14797). AT and OJ were supported by the European Union
through the European Social Fund (Mobilitas Grants no. MTT77 and MJD447). SL
acknowledges the funding from the Austrian Fonds zur F\"orderung der
wissenschaftlichen Forschung (FWF) via a Schr\"odinger fellowship J3247-N16.


%

\clearpage

\begin{table*}[h]
\begin{tabular}{c}
\huge{\texttt{Supporting Material}} \\
\end{tabular}
\end{table*}

\begin{figure*}[!h]
\includegraphics[width=11cm]{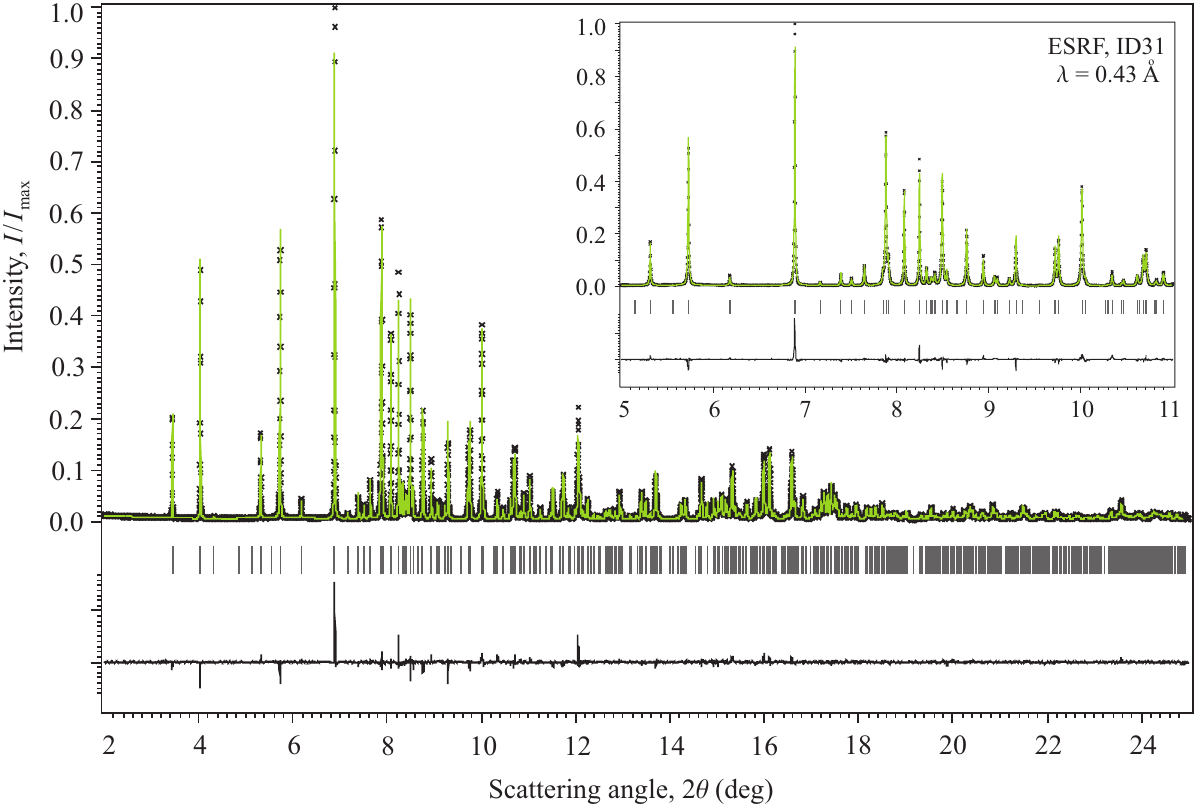}
\begin{minipage}{14cm}
\caption{\label{fig:s1}\normalsize
Rietveld refinement of the room-temperature high-resolution XRD data. Ticks show the reflection positions of clinoclase.
}
\end{minipage}
\end{figure*}

\begin{table*}[!hb]
\begin{minipage}{14cm}
\caption{
The lattice parameters, atomic positions and atomic displacements parameters $U_{\iso}$ (in 10$^{-2}$~\r A$^2$) refined from the high-resolution powder XRD data on the natural sample of clinoclase. Positions of hydrogen atoms are determined by a DFT-based structure optimization within GGA. All atoms are in the $4e$ Wyckoff position of the $P2_1/c$ space group.
}
\begin{ruledtabular}
\begin{tabular}{lcccc}
  $a=7.2658$~\r A & $b=6.4588$~\r A & $c=12.3929$~\r A & $\beta=99.487^{\circ}$ \\
\\
  Atom & $x/a$     & $y/b$     & $z/c$     & $U_{\iso}$ \\
  Cu1  & 0.7864(2) & 0.1400(2) & 0.3296(1) &  0.74(3)   \\
  Cu2  & 0.8141(2) & 0.3802(2) & 0.1272(1) &  0.77(3)   \\
  Cu3  & 0.3856(2) & 0.3527(2) & 0.4120(1) &  0.99(4)   \\
  As   & 0.3083(1) & 0.1509(2) & 0.1794(1) &  0.85(3)   \\
  O1   & 0.4090(7) & 0.0712(8) & 0.0714(4) &  0.14(6)$^a$ \\
  O2   & 0.8387(7) & 0.8447(8) & 0.3681(4) &  0.14(6)$^a$ \\
  O3   & 0.1802(7) & 0.9408(8) & 0.2108(4) &  0.14(6)$^a$ \\
  O4   & 0.4705(8) & 0.2147(7) & 0.2813(4) &  0.14(6)$^a$ \\
  O5   & 0.7777(7) & 0.2039(7) & 0.4780(4) &  0.14(6)$^a$ \\
  O6   & 0.8047(7) & 0.0957(8) & 0.1747(4) &  0.14(6)$^a$ \\
  O7   & 0.1871(7) & 0.1705(8) & 0.4157(4) &  0.14(6)$^a$ \\
  H1$^b$ & 0.7493  & 0.3515    & 0.4807    &    --        \\
  H2$^b$ & 0.9346  & 0.4680    & 0.6787    &    --        \\
  H3$^b$ & 0.1511  & 0.1648    & 0.4864    &    --        \\
\end{tabular}
\end{ruledtabular}
\begin{flushleft}
$^a$ The atomic displacements parameters of all oxygen atoms were refined as a single parameter

$^b$ Hydrogen positions are obtained from the structure relaxation within GGA
\end{flushleft}
\end{minipage}
\end{table*}

\begin{table*}[tbp]
\begin{ruledtabular} 
\caption{\label{H_rot} 
The $t_{\text{D1}}$ and $t$ hopping parameters (in meV) as function of the out-of-plane angle of H, $\tau$. $U_{\text{eff}}$\,=\,4.0\,eV in the AFM exchange contributions $J^{\text{AFM}}_{\text{D1}}$ (in K). The ferromagnetic contribution is defined as $J^{\text{FM}}_{\text{D1}}=J_{\text{D1}}-J^{\text{AFM}}_{\text{D1}}$. $J_{\text{D1}}$ (in K) is calculated with the Cu$_2$O$_6$H$_5$ cluster model and a PBE0 hybrid functional.
}
\begin{tabular}{p{0.8cm} p{0.8cm} p{0.8cm} p{0.8cm} p{0.8cm} p{0.8cm}}
$\tau$ & $t_{\text{D1}}$ & $t$  & $J^{\text{AFM}}_{\text{D1}}$  & $J^{\text{FM}}_{\text{D1}}$ &  $J_{\text{D1}}$  \\ \hline
   7   &   -186          & 292  &    401                        & 213                         & 188 \\
  11   &   -184          & 289  &    393                        & 208                         & 185 \\
  22   &   -169          & 283  &    331                        & 174                         & 157 \\
  31   &   -155          & 281  &    279                        & 161                         & 118 \\
  39   &   -144          & 278  &    241                        & 163                         &  78  \\
  52   &   -128          & 276  &    190                        & 175                         &  15  \\
  67   &   -115          & 276  &    153                        & 183                         & -30 \\
\end{tabular}
\end{ruledtabular}
\end{table*}

\begin{figure*}[tbp]
\includegraphics[width=8cm]{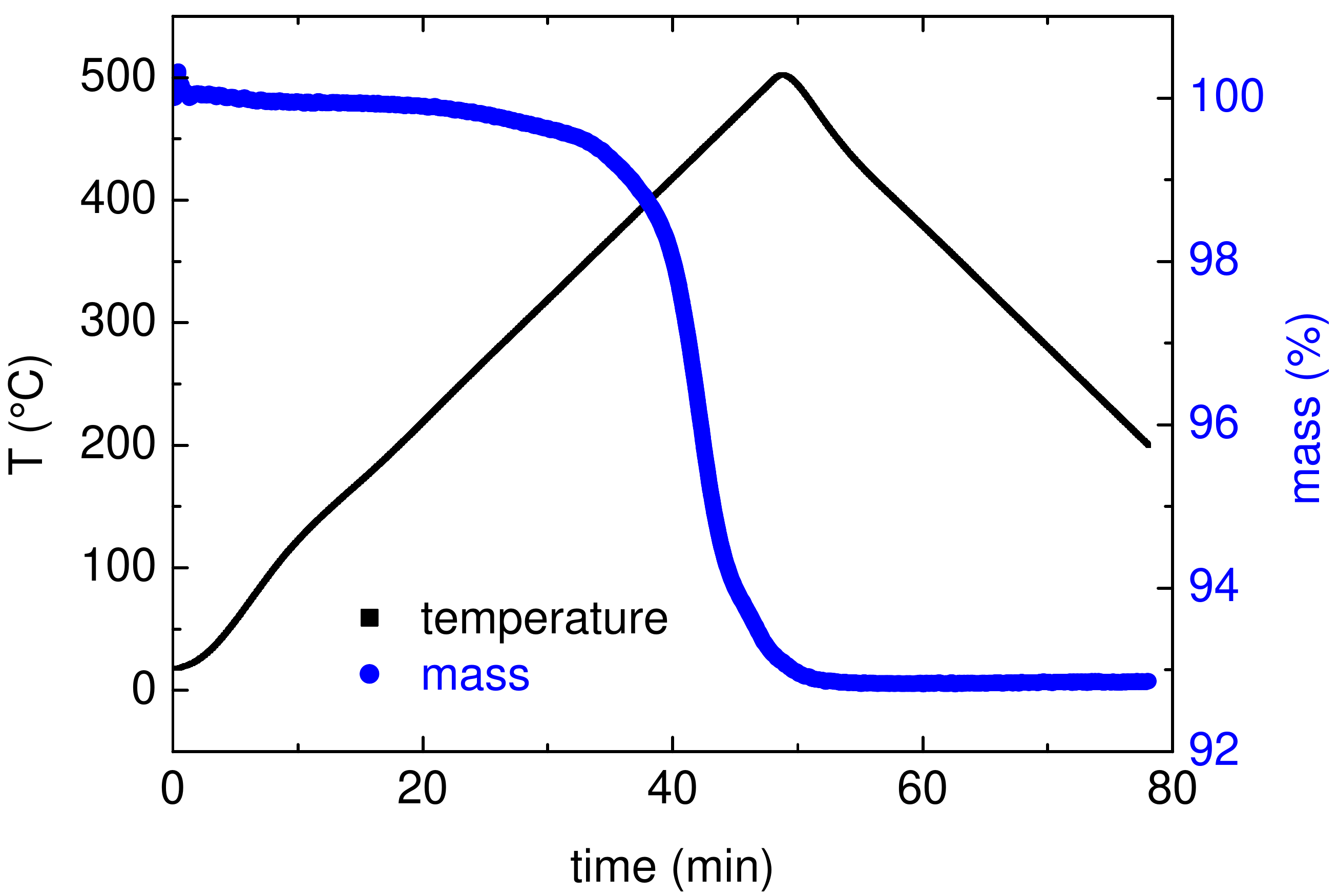}
\caption{\label{TG}
(Color online) Thermogravimetric analysis data taken in a temperature range of 20--500~$^{\circ}$C. The mass loss amounts to 7.13\%.}
\end{figure*}

\begin{figure*}[tbp]
\includegraphics[width=8cm]{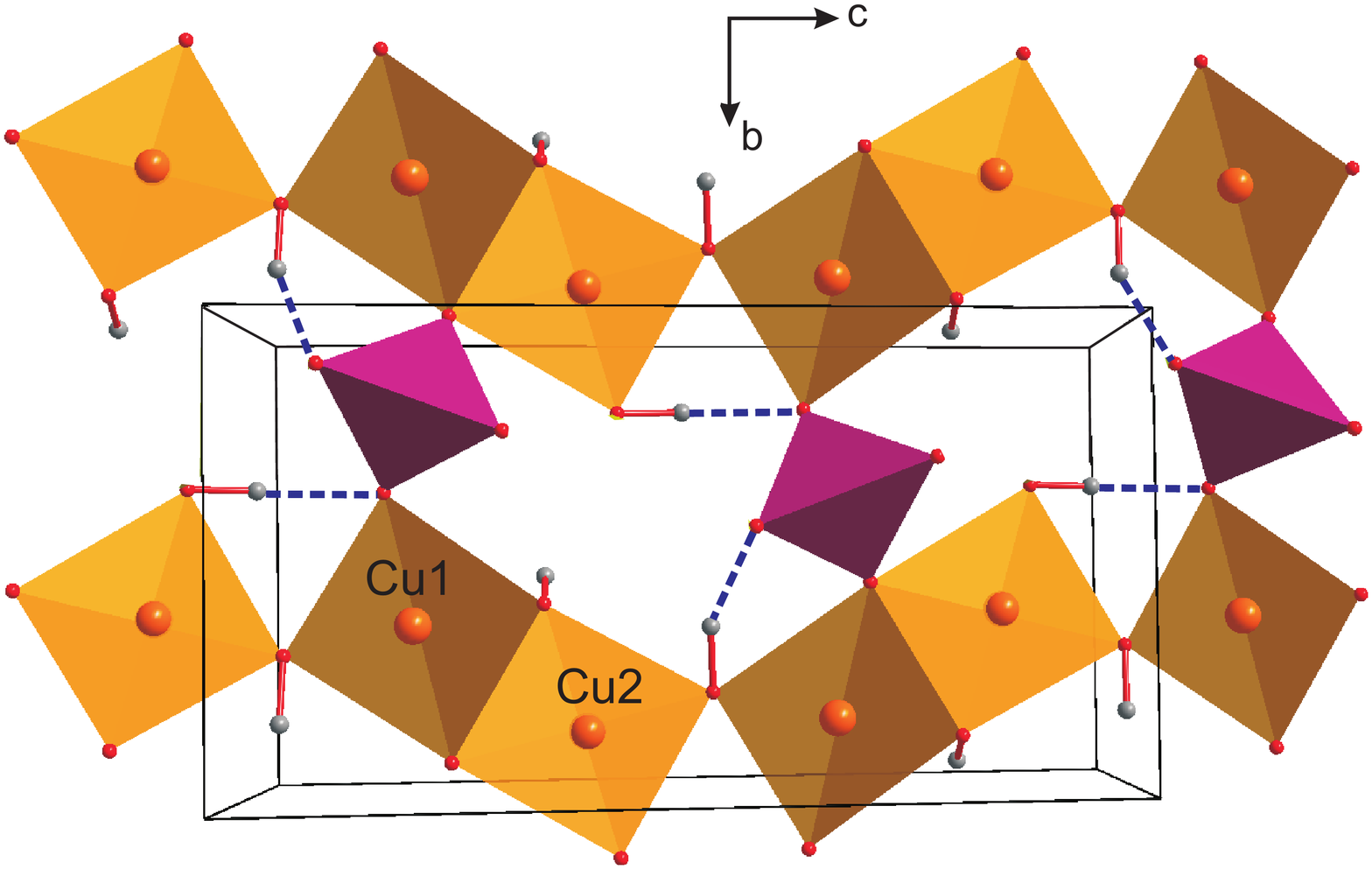}
\caption{\label{Hbond}
(Color online) The hydrogen bonds (blue) within the $bc$-plane of the clinoclase structure. The D1 dimers are shown in orange and the AsO$_4$ tetrahedra in pink color.  }
\end{figure*}

\end{document}